\documentclass[aps,pra,preprint,groupedaddress,showkeys,floatfix]{revtex4}
\usepackage[dvips]{graphicx}

\begin{document}
\title{Polarization Effects in Two-Photon Free–Free Transitions in Laser-Assisted Electron–Hydrogen Collisions}

\author{Aurelia Cionga and Gabriela Buica}
\affiliation{Institute for Space Sciences, P.O. Box MG-36, Bucharest-M\u agurele,Bucharest, R-76900 Romania}
%{\rm e-mail: cionga@venus.ifa.ro}}

\begin{abstract}
Two-photon free-free transitions in elastic laser-assisted
electron-hydrogen collisions are studied in the domain of high scattering
energies and low or moderate field intensities, in the third order of
perturbation theory, taking into account all the involved Feynman diagrams.
Based on the analytical expressions of the transition amplitudes,
the differential cross sections for two-photon absorption/emission
are computed at impact energy $E_i=100$ eV.
The effect of field polarizations on the angular distribution
and on the frequency dependence of the differential cross section
is analyzed.
\end{abstract}
\maketitle

\section{Introduction}

Recently, a series of experimental \cite{exp}
and theoretical \cite{teo} works have been devoted to the
study of free-free transitions in laser-assisted elastic electron-atom
collisions at low scattering energies.

It is the aim of this work to investigate free-free transitions
in a different regime, that of high scattering energies and low
or moderate field intensities such that the use of perturbation
theory might provide a sensible description of the process.
We focus our attention on the study of
free-free transitions that involve the absorption/emission of two
{\it different} photons by the compound projectile-target
system in an external radiation field.
The target is the hydrogen atom in the ground state.
The process can be formally represented by
\begin{eqnarray}
H({\rm E}_{1s})  +  e^-\left({\vec k}_i, E_i\right)
	\pm \left[ \gamma \left({\vec \varepsilon_1}, \omega_1 \right)
		  +\gamma \left({\vec \varepsilon_2}, \omega_2 \right)\right]
	\to
        %\nonumber \\
H({\rm E}_{1s})  +  e^-\left({\vec k}_f, E_f\right)
,\label{proc}
\end{eqnarray}
where
$E_{i(f)}$, $\vec k_{i(f)}$ are the initial (final) energy and momentum
of the projectile;
$\omega_j$, $\vec \varepsilon_j$ denote the frequency and the polarization
vector of the photon $j$ ($j=1,2$).
The upper sign corresponds to the absorption
of both photons, the lower one corresponds to their stimulated emission.

The process (\ref{proc}) has been previously investigated for two
{\it identical} photons.
Kracke {\it et al} \cite{bri} have studied the differential cross section of
two-photon free-free transitions at high scattering energies (50-500 eV)
for photon energies below the ionization threshold of hydrogen
($\omega < 20$ eV). They have used in their work the lowest order
perturbation theory, taken into account all the involved Feynman diagrams.
For strong fields, the laser-particle interaction must be treated
beyond the perturbation theory.
In this context, D{\"o}rr {\it et al} \cite{mar} developed the
%so called
Born-Floquet theory, in which both laser-projectile and laser-target
interactions are treated exactly. This approach is valid in
the domain of high scattering energies
since it involves the first Born
approximation to treat the projectile-target interaction.
We refer to this paper for a comprehensive analysis of other previous works.

In Sec. II we present the formalism we have used to evaluate transition
matrix elements for two-photon absorption/emission: the projectile-target
interaction as well as the interaction between the electrons and the
electromagnetic field have been treated perturbatively.
We have evaluated the analytic expressions of the corresponding
transition amplitudes
in the third order of perturbation theory,
including the twenty-four Feynman diagrams.
The third section is
devoted to the discussion of the numerical results. We report here
our results concerning the influence of the state of polarization
of the two photons on the differential cross section of the scattered
electron. We claim that this effect is significant in the domain in
which the dressing of the target is important:
at small scattering angles in general and
in particular close to atomic resonances.

\section{BASIC EQUATIONS}

The time evolution of the electron-hydrogen system in the presence of
an electromagnetic field described by the vector potential
\begin{equation}
\vec{\cal A}(t)=\vec \varepsilon_1 A_{01} \cos(\omega_1t) +
		\vec \varepsilon_2 A_{02} \cos(\omega_2t)
,\label{field}
\end{equation}
is governed by the hamiltonian
\begin{eqnarray}
{\cal H}&  = &
	\frac {\vec{p}^2}{2} -\frac{1}{r} + \frac{\vec{P}^2} 2
      %  &\nonumber\\ & + &
           +     \frac{1} {\vert\vec{r} - \vec{R}\vert} - \frac{1}{R} +
		\frac 1 c \left[ \vec{p} +
                \vec{P} \right] \cdot \vec{\cal A}(t) \nonumber \\
         &\equiv & H_0 + V + W(t) ,
\label{ha}
\end{eqnarray}
where
$\vec{r}$, $\vec{p}$ are the position and momentum operator
of the bound (atomic) electron
and
$\vec{R}$, $\vec{P}$ are the position and momentum operator
of the free (projectile) electron.
$V \equiv -R^{-1} + \vert\vec{r} - \vec{R}\vert^{-1}$
denotes the e-H interaction in the direct channel
and $W(t) \equiv c^{-1} \left[ \vec{p} +
\vec{P} \right] \cdot \vec{\cal A}(t)$
denotes the interaction of the charge particles with the
field, treated in the velocity gauge, using the dipole approximation.
The $\vec {\cal A}^2$-term was eliminated through a unitary transformation.

In the {\it first nonvanishing order} of the perturbation theory,
the $S-$matrix elements corresponding to two-photon processes are given by
\begin{equation}
S^{(2)} = -
%(-i)^2
\int_{-\infty}^{+\infty}dt_1\int_{-\infty}^{t_1}dt_2 <{\chi}%
_f^- |\tilde W(t_1) \tilde W(t_2) | {\chi}_i^+ >
,\label{smat}
\end{equation}
where $\tilde W(t) = e^{iH_0t}W(t)e^{-iH_0t}$.
In the previous equation
$|\chi _i^{+}>$ and $|\chi _f^{-}>$ describe the initial and final states
of the colliding system (electron-atom)
\begin{eqnarray}
|\chi _i^{+}>& = &|\Psi _i>+G^{+}({\cal {E}}_i)V|\Psi _i>,\\
|\chi _f^{-}>& = &|\Psi _f>+G^{-}({\cal {E}}_f)V|\Psi _f>,
\end{eqnarray}
where
\begin{equation}
G^{\pm} ({\cal E}) = \left[{\cal E}-H_0-V\pm i\delta \right]^{-1}
\end{equation}
and $\delta$ a positive infinitesimal number.
$|\Psi _{i,f}>$ are the asymptotic states corresponding to the
colliding system in the absence of the interaction~$V$
\begin{eqnarray}
|\Psi _i>&=&|\psi _{1s}>|K_i>, \\
|\Psi _f>&=&|\psi _{1s}>|K_f>.
\end{eqnarray}
Here $|\psi_{1s}> $ denotes the ground state of a hydrogen atom
%$H_A|\psi _{i,f}>=E_{i,f}|\psi _{i,f}>,$
and $|K_{i,f}>$ are plane waves.
The initial and final energies of the electron-atom system are
\begin{eqnarray}
{\cal {E}}_i&=&{\rm E}_{1s}+\frac{k_i^2}{2},\\
{\cal {E}}_f&=&{\rm E}_{1s}+\frac{k_f^2}{2}\pm \left(\omega_1+\omega_2 \right)
.\end{eqnarray}

The transition-matrix element involving two {\it different} photons,
both absorbed or emitted, is given by
\begin{eqnarray}
T^{(2)} = \frac{1}{4}
        (1+{\cal P}_{12})
%        \\        &\times &<
     {\chi}_f^- | {\vec A}_1\cdot(\vec p+\vec{P}) G^+ ({\cal {E}}_i
     \pm \omega_2) {\vec A}_2\cdot(\vec p+\vec{P}) | {\chi}_i^+ >
, \label{tt}
\end{eqnarray}
where ${\cal P}_{12}$ is the permutation operator between
the vector potentials ${\vec A}_j= {\vec \varepsilon}_j A_{0j}$ (j=1,2),
which describe the two components of the field (\ref{field}).
In Eq.(\ref{tt}) the upper sign corresponds to absorption, the lower one to
stimulated emission.
It is possible to write this matrix element as the sum of three terms,
each of them connected with specific Feynman diagrams, as we shall discuss
later in this section. These terms are:\\
{\it - the electronic term}
\begin{eqnarray}
T_P =\frac{1}{4}(1+{\cal P}_{12})
%\\    &\times &
<{\chi }_f^{-}|\vec{A}_1\cdot \vec{P}G^{+}({\cal {E}}_i \pm \omega_2)
        \vec{A}_2\cdot \vec{P}|{\chi }_i^{+}>
,\label{te}
\end{eqnarray}
{\it - the mixed term}
\begin{eqnarray}
T_M =\frac{1}{4}(1+{\cal P}_{12}) (1+{\cal P}_{\vec p \vec P})
%\\    &\times &
<{\chi }_f^{-}|\vec{A}_1\cdot \vec{p}G^{+}({\cal {E}}_i
     \pm \omega_2)\vec{A}_2\cdot \vec{P}|{\chi }_i^{+}>
,\label{tm}
\end{eqnarray}
where ${\cal P}_{\vec p \vec P}$ is the permutation operator between
$\vec{p}$ and $\vec{P}$, and  \\
{\it - the atomic term}
\begin{eqnarray}
T_A = \frac{1}{4}(1+{\cal P}_{12})
% \\      &\times &
    <{\chi }_f^{-}|\vec{A}_1\cdot \vec{p}G^{+}({\cal {E}}_i
   \pm \omega_2 )\vec{A}_2\cdot \vec{p}|{\chi }_i^{+}>
.\label{ta}
\end{eqnarray}
Since we restrict ourselves to the domain of high scattering energies,
we use the first Born approximation to treat electron-atom scattering,
which implies
\begin{eqnarray}
|\chi _i^{+}> & \simeq & |\Psi _i>+G_0^{+}({\cal {E}}_i)V|\Psi _i>, \\
|\chi _f^{-}> & \simeq & |\Psi _f>+G_0^{-}({\cal {E}}_f)V|\Psi _f>,
\end{eqnarray}
where
\begin{equation}
G_0^{\pm }({\cal {E}})=[{\cal {E}}-H_0\pm i\delta ]^{-1}.
\end{equation}
In this way, the evaluation of the transition matrix element
is made in the third order of perturbation theory:
the second order in the electric field and
the first order in the scattering potential, $V$.

\subsection{Electronic term}

The electronic term is connected to six Feynman diagrams
in which only the projectile exchanges two different photons
with the field (\ref{field}).
Only three of these diagrams
are represented in Fig.1(a), the other three are obtained by interchanging
$\omega_1$ and $\omega_2$.

In the standard way,
after integration over the projectile coordinates, the electronic term
in Eq.(\ref{te}) may be written as
\begin{equation}
T_P=\frac{\sqrt{I_1 I_2}}{4 \omega_1^2 \omega_2^2}
         (\vec{\varepsilon_1}\cdot\vec{q})(\vec{\varepsilon_2}\cdot\vec{q})
             <\psi_{1s}|F(\vec q)|\psi_{1s}>
,\label{te1}
\end{equation}
where $I_j$ is the intensity of the component $j$ of the field (\ref{field}),
$\vec q$ is the momentum transfer of the projectile and
$F(\vec{q})$ is the form factor operator
$$
F(\vec{q})=\frac 1{2\pi ^2q^2}\left[ \exp {(i\vec{q}\cdot \vec{r})}-1\right]
.$$

We remind that this is the only term which gives contributions to the
weak field intensity limit of Bunkin-Fedorov formula \cite{b-f}.
That approach describes the target by a potential
and neglects the atomic dressing. In order to take into account the atomic
dressing we include in our calculation the other eighteen Feynman diagrams,
corresponding to the mixed and atomic terms.

\subsection{Mixed term}

The mixed term is connected to twelve Feynman diagrams, in which each electron
(free and bound) absorbs/emits one photon from each component of
the field (\ref{field}).
Only six diagrams
are represented in Fig.1(b), the other six are obtained again by
interchanging $\omega_1$ and $\omega_2$.

In order to evaluate the mixed term
we took advantage of the analytic form of the vectors
$$\vert{\vec w}_{100}(\Omega)>=
- G_C(\Omega)\vec{p} \vert \psi_{1s}>,$$
which were previously studied in Ref.\cite{vf1}.
Here $G_C(\Omega) $ is the Coulomb Green function.
After integration over the projectile coordinates, the mixed term in
Eq.(\ref{tm}) is written as
\begin{eqnarray}
T_M &=&\mp \frac{\sqrt{I_1 I_2}}{4 \omega_1 \omega_2}
%&\times &
       \left\{ \right.\frac{\vec{\varepsilon_2}\cdot\vec{q}}{\omega_2}
       \left[<\psi_{1s}|F(\vec q)|
       \vec{\varepsilon_1}\cdot {\vec{w}}_{100}(\Omega_{1}^{\pm})>
%        \right.
 %&&\hspace*{0.9cm}+\left.
        <\vec{\varepsilon_1}\cdot {\vec{w}}_{100}(\Omega_{1}^{\mp})|
                 F(\vec q)|\psi_{1s}>\right]
        \label{tm1}\\
&& \left. \hspace*{1.4cm}+
  \frac{\vec{\varepsilon_1}\cdot\vec{q}}{\omega_1}
\left[<\psi_{1s}|F(\vec q)|
         \vec{\varepsilon_2}\cdot {\vec{w}}_{100}(\Omega_{2}^{\pm})>
%\right. \nonumber \\
%&&\hspace*{0.9cm}+\left. \left.
        <\vec{\varepsilon_2}\cdot {\vec{w}}_{100}(\Omega_{2}^{\mp})|
          F(\vec q)|\psi_{1s}>\right]\right\}
,\nonumber
\end{eqnarray}
where the parameters
$\Omega^{\pm}_{1,2}$ are given by
\[
\Omega_{1,2}^{\pm}= {\rm E}_{1s} \pm\omega_{1,2}
.\label{par1}
\]
In Eq.(\ref{tm1}) the upper signs correspond to absorption and the
lower ones correspond to stimulated emission of both photons.

The atomic matrix elements in Eq.(\ref{tm1}), which appear also in
one photon processes, have been evaluated
analytically \cite{ac1}. Based on this result it is possible to write down
the general structure of the mixed term as
\begin{equation}
T_M  =\sqrt{I_1 I_2} (\vec {\varepsilon}_1\cdot\vec{q})
			(\vec {\varepsilon}_2 \cdot \vec{q})
	{\cal T_M} \left( \omega_1, \omega_2; q \right)
,\label{tm2}
\end{equation}
where the radial part, ${\cal T_M} \left( \omega_1, \omega_2; q \right)$,
is expressed in terms of hypergeometric functions.

\subsection{Atomic term}

The atomic term is connected to six Feynman diagrams in which two different photons
are exchanged between the bound electron and the field (\ref{field}).
Only three diagrams
are represented in Fig.1(c), the other three are obtained by interchanging
$\omega_1$ and $\omega_2$.

Our analytic formula for the atomic term is computed using the tensors
%in Eq.(\ref{ta})
$$| w_{ij,100} (\Omega^{\prime}, \Omega ) > = G_C( \Omega^{\prime} ) p_i G_C
( \Omega ) p_j | \psi_{1s} >,$$
studied in Ref.\cite{vf2}.
After integration over the coordinates of the projectile, the atomic term in
Eq.(\ref{ta}) is written as
\begin{eqnarray}
T_A &=&\frac{\sqrt{I_1 I_2}}{4 \omega_1 \omega_2}
       \sum_{j,l=1}^3 \varepsilon_{1j}\varepsilon_{2l}
       %\\
%&\times & \;
\left[ \;\;
    <w_{j,100}(\Omega_{1}^{\mp})|F(\vec q)|{w}_{l,100}(\Omega_{2}^{\pm})>
%        \right.
%\nonumber \\
%&&\hspace*{0.05cm} +
    <{w}_{l,100}(\Omega_{2}^{\mp})|F(\vec q)|w_{j,100}(\Omega_{1}^{\pm})>
      \right.   \nonumber\\
&&\hspace*{2.9cm}+
    <\psi_{1s}|F(\vec q)| w_{jl,100} (\Omega^{'\pm}, \Omega_{2}^{\pm}) >
 %       \nonumber \\
%&& \hspace*{0.05cm} +
    <\psi_{1s}|F(\vec q)| w_{lj,100} (\Omega^{'\pm}, \Omega_{1}^{\pm}) >
 	\nonumber\\
&&\left.\hspace*{2.9cm} +
    <w_{lj,100} (\Omega^{'\mp}, \Omega_{1}^{\mp})|F(\vec q)|\psi_{1s}>
%        \nonumber \\
%&&\hspace*{0.05cm} +\left.
    <w_{jl,100} (\Omega^{'\mp}, \Omega_{2}^{\mp} )|F(\vec q)|\psi_{1s}>\;
 	\right]
.\nonumber\\
  \label{ta1}
\end{eqnarray}
Here the parameter $\Omega^{\prime}$ takes the values
\[
\Omega^{\prime \pm}= {\rm E}_{1s} \pm(\omega_1+\omega_2)
\label{par2}
\]
and $\Omega^{\pm}_{1,2}$ were defined above.
In Eq.(\ref{ta1}) the upper signs correspond to absorption
and the lower ones correspond to emission of two different photons.

We have evaluated analytically the atomic matrix element in Eq.(\ref{ta1});
based on this results one can write the general structure of the atomic term as
\begin{eqnarray}
T_A = \sqrt{I_1 I_2}
%&
\left[
%&\right.&
	(\vec {\varepsilon}_1\cdot\vec{q})
		(\vec {\varepsilon}_2 \cdot \vec{q})
	{\cal T}^{\prime}_{\cal A} \left( \omega_1, \omega_2; q \right)
%        \nonumber\\
%&& + \left.
	(\vec {\varepsilon}_1\cdot\vec {\varepsilon}_2)
{\cal T}^{\prime \prime}_{\cal A} \left( \omega_1, \omega_2; q \right)
	\right]
,\label{ta2}
\end{eqnarray}
where the radial parts,
${\cal T}^{\prime}_{\cal A}\left( \omega_1, \omega_2; q \right)$ and
${\cal T}^{\prime \prime}_{\cal A}\left( \omega_1, \omega_2; q \right)$,
are expressed as series of hypergeometric functions \cite{a-g}.

For the sake of simplicity, the equations (\ref{te1}-\ref{ta2})
have been written using linear polarizations.
We point out that for photon emission,
one must take the complex conjugate of the polarization vector.

Finally, the differential cross section for the absorption/emission
of two different photons in laser-assisted elastic electron-hydrogen collisions
can be written as
\begin{equation}
\frac{d\sigma}{d\Omega} = {(2\pi)}^4 \frac{k_f}{k_i} {|T_P+T_M+T_A |}^2
,\label{sec}
\end{equation}
where the electronic, mixed, and atomic terms have the structure given in
Eqs.(\ref{te1}, \ref{tm2}, \ref{ta2}).

\section{RESULTS}

We have computed the differential cross section for two-photon
free-free transitions in laser-assisted elastic electron-hydrogen collisions
at scattering energy $E_i=100$ eV.
We have chosen to report
here the case of two laser sources having
the same frequency,
$\omega_1=\omega_2 \equiv \omega$,
but different polarizations.
The investigated photon energies are smaller than the ionization energy
of hydrogen. The results are valid for low and moderate field intensities,
bellow 10$^{10}$ W/cm$^2$.
In all the cases that we have studied the initial momentum of the
projectile, ${\vec k}_i$, defines the $Oz$-axis.

We discuss here the effect of the state of polarization of the
photons on the frequency dependence of the differential cross section
and on the azimuthal angular distribution of the scattered electrons,
for scattering angles in the domain where
target dressing effects are important. In general, this domain corresponds
to small scattering angles, as it has been pointed out
by Kracke {\it et al} \cite{bri}, who studied the monochromatic case.

\subsection{Frequency dependence}

In Fig.2 we present the differential cross section for two-photon absorption
in Eq.(\ref{sec}), normalized with respect to the field intensities, $I_1I_2$,
as a function of the photon frequency, in the range $0<\omega<6.8$ eV.
The scattering angle, $\theta=5^0$, is in the domain where the
dressing effects are important.
Our calculations were performed for linear polarizations. We have chosen
${\vec \varepsilon}_1 || Oz $ and
${\vec \varepsilon}_2 \perp \vec \varepsilon_1$,
the polarization vectors defining the scattering plane.

The differential cross section (solid line) exhibits a
series of resonances, located between 6 and 6.8 eV.
These resonances occur at photon frequencies such that
$2\omega= |{\rm E}_{1s}| (1-1/n^2)$ for $n>2$, where $n$ is
the principal quantum number.
They correspond to poles in the radial integrals
${\cal T}^{\prime}_{\cal A}$ and ${\cal T}^{\prime \prime}_{\cal A}$,
which appear in the atomic term.
%(\ref{ta2}).
The resonance corresponding to $n$=2, i.e., $\omega=5$~eV,
%$2\omega$=3$|{\rm E}_{1s}|/4$,
does not exist for orthogonal polarization:
the connected pole appears in Eq.(\ref{ta2}) only in the radial integral
${\cal T}^{\prime \prime}_{\cal A}$,
which multiplies the scalar product
$\vec \varepsilon_1 \cdot \vec \varepsilon_2$, which vanishes.
To emphasize the origin of these resonances we have plotted also in Fig.2 the
electronic (dotted) and mixed (dot-dashed) contributions, which were
calculated when only $T_P$ and $T_M$, respectively, were taken into
account in Eq.(\ref{sec}).

We believe that this series of resonances is particularly interesting
from the experimental point of view. Indeed, two-photon processes may be
easier to detect at photon energies close to one of these resonances because
they do not correspond to resonances of the lower order processes,
namely one-photon absorption/emission.

A second series of resonances, not shown in Fig.2,
is located between 10 and 13.6 eV.
They occur for photon frequencies such that
$\omega=|{\rm E}_{1s}| (1-1/n^2)$, where $n\ge2$.
This time the resonances correspond to poles which exist in three
radial integrals: ${\cal T}_{\cal M}$, ${\cal T}^{\prime}_{\cal A}$, and
${\cal T}^{\prime \prime}_{\cal A}$.

In Fig.2 the differential cross section (solid line) has also
a series of minima. The first minimum
%($\omega \simeq 0.4$ eV)
is due to the fact that, in this geometry, the differential
cross section in Eq.(\ref{sec}) is proportional to
$|({\vec \varepsilon}_1 \cdot {\vec q})|^2 =
\mid k_i - k_f \cos \theta \mid^2$,
which is vanishing at $\theta$ = 5$^0$.
The other minima are due to interferences between the electronic, mixed,
and atomic terms.

\subsection{Azimuthal angular distribution}

We have found out that the azimuthal angular distributions of the scattered
electrons are significantly modified in the case of
complex polarization vectors if virtual transitions to continuum
are energetically allowed, i.e., $2\omega>|{\rm E}_{1s}|$.
To illustrate this remark we discuss two distinct cases.
In the first case the photon frequency corresponds to KrF laser,
$\omega=5$ eV. One has $2\omega < | {\rm E}_{1s} |$ and the radial integrals
${\cal T}_{\cal M}$, ${\cal T}^{\prime}_{\cal A}$, and
${\cal T}^{\prime \prime}_{\cal A}$ are real.
On the contrary, in the second case, when the frequency of the photons
corresponds to the second harmonic of KrF, one has $2\omega > | {\rm E}_{1s} |$
and the corresponding radial integrals are complex.
Each of these frequencies corresponds to atomic resonances.
For each of these frequencies,
we present the differential cross section of two-photon absorption
%in Eq.(\ref{sec})
at a fixed scattering angle, $\theta$ = 20$^0$,
as a function of the azimuthal angle $\phi$.
Two different choices of the polarization vectors were investigated.

\vspace*{0.5cm}

1.\hspace*{0.5cm}
${\vec \varepsilon_1} = {\vec e}_z$ and
${\vec \varepsilon_2} = \left({\vec e}_z +i {\vec e}_x\right)/\sqrt 2$

\vspace*{0.1cm}

In this case one laser beam, which direction defines the $Ox$ axis,
is linearly polarized with the polarization vector parallel to the initial
momentum of the projectile, $\vec \varepsilon_1 || Oz$.
The other laser beam, of the same frequency, defines the $Oy$-axis and is
circularly polarized,
${\vec \varepsilon_2} \equiv \left({\vec e}_z +i {\vec e}_x\right)/\sqrt 2$.

For this choice of polarizations
the electronic and the mixed terms in Eqs.(\ref{te1},\ref{tm2}) have the same
$\phi$-dependence, namely of the form
$\alpha (\alpha  + i \beta \cos \phi)$.
%{\cal T}_{\cal M}
As a consequence, the electronic and the mixed contribution
to the differential cross section
are symmetric with respect to the reflection in the planes $xOz$ and $yOz$
for both frequencies, $\omega=5$eV in Fig.3 and $\omega=10$eV in Fig.4.
%$\omega$ and $\omega^{\prime}$.

The atomic term (\ref{ta2}) has a different $\phi$-dependence, given by

\begin{equation}
T_{A} \sim {\cal T}^{\prime \prime}_{\cal A}
	+ \alpha^2 {\cal T}^{\prime}_{\cal A}
+ i \alpha \beta {\cal T}^{\prime}_{\cal A}\cos \phi
.\label{pol11}
\end{equation}
The atomic contribution
has the above mentioned reflection properties only in Fig.3,
where $\omega$ = 5 eV. In Fig.4, where $\omega $ = 10 eV,
the atomic contribution has only one symmetry plane.
%namely $xOz$ (Fig.4).
Indeed, $\phi \to -\phi$ is a symmetry operation for
the quantity in Eq.(\ref{pol11}), therefore $xOz$ is a symmetry plane.
$yOz$ is no more a symmetry plane because
both ${\cal T}^{\prime}_{\cal A}$ and
${\cal T}^{\prime\prime}_{\cal A}$
are complex and the modulus square of the quantity in Eq.(\ref{pol11})
is not symmetric to the change $\phi=\pi/2-\xi \to \pi/2+\xi$.

At small scattering angles in general and in particular close
to resonances, where the dressing effects are important, there are
interferences between the atomic and the mixed
contributions. They impose the $\phi-$dependence of the
differential cross section (\ref{sec}), given by
\begin{equation}
\frac{d\sigma}{d\Omega} \sim
%|T_P+T_M+T_A|^2 \simeq
|{\cal T}^{\prime \prime}_{\cal A} + \alpha^2 {\cal T}_1+
	i \alpha \beta {\cal T}_1\cos \phi |^2
,\label{pol12}
\end{equation}
where
${\cal T}_1 ={\cal T}_{\cal P}+{\cal T}_{\cal M}+{\cal T}_{\cal A}^{\prime}$
and ${\cal T}_{\cal P}$ is the radial part of the electronic term (\ref{te1}).
In particular, only ${\cal T}^{\prime\prime}_{\cal A}$ has a pole
when $\omega$ = 5 eV, therefore the atomic contribution in Fig.3
is almost $\phi-$independent (a circle) and it is dominant in the differential cross section.
The situation is different in Fig.4 because
three radial integrals, ${\cal T}_{\cal M}$, ${\cal T}^{\prime}_{\cal A}$, and
${\cal T}^{\prime\prime}_{\cal A}$,
have poles when $\omega $ = 10 eV .

At large scattering angles the electronic term is dominant and it imposes the
angular distribution of the differential cross section.

We note also that
the change from right to left circular polarization
%${\vec \varepsilon}_2 \to {\vec \varepsilon}_2^*$,
%${\vec \varepsilon_2} \equiv \left({\vec e}_z - i {\vec e}_x\right)/\sqrt 2$
implies a simultaneous change of the sign of the last term
in Eqs.(\ref{pol11}-\ref{pol12}), which is equivalent to a rotation by $\pi$
%$\phi \to \phi+\pi$,
of the curves in Figs.3-4. This rotation is visible in the angular
distributions only if the radial integrals have complex values,
i.e., when $2\omega > |E_{1s}|$.

\vspace*{0.5cm}

2.\hspace*{0.5cm}
${\vec \varepsilon_1} = \left({\vec e}_z +i {\vec e}_x\right)/\sqrt 2$
and
${\vec \varepsilon_2} = \left({\vec e}_y +i {\vec e}_z\right)/\sqrt 2$

\vspace*{0.1cm}

In this second case both polarization vectors are right circularly polarized.
The direction of the first laser beam defines the $Oy$-axis,
the other one the $Ox$-axis.

The electronic and the mixed terms have again the same
$\phi$-dependence, given by
$$ \alpha \beta (\sin \phi - \cos \phi) +
i(\alpha^2 + \beta^2 \sin \phi \cos \phi).$$
For both frequencies, $\omega =5$ eV in Fig.5 and $\omega =10$~eV in Fig.6,
the first ($\phi=\pi/4$) and the second ($\phi=3\pi/4$)
bisector of the angle $xOy$ are symmetry axes of
the electronic and the mixed contributions to the differential cross section.

The atomic term (\ref{ta2}) has a different $\phi$-dependence, namely
\begin{eqnarray}
T_{A} &\sim& \alpha \beta {\cal T}^{\prime}_{\cal A}(\sin \phi - \cos \phi)
\label{pol22} \\
&+& i\left({\cal T}^{\prime \prime}_{\cal A}
	+ \alpha^2 {\cal T}^{\prime}_{\cal A}
		+ \beta^2 {\cal T}^{\prime}_{\cal A} \sin \phi \cos \phi \right)
.\nonumber
\end{eqnarray}
The atomic contribution
has the two previous symmetry axes only when $\omega$ = 5 eV (Fig.5);
when $\omega$ = 10~eV
the atomic part has only one symmetry axis: along the second bisector (Fig.6).
Since the atomic term is always important for the chosen value of $\theta$,
it will determine the symmetry
properties of the differential cross section.

The study of two photon with left circular polarizations
implies the change of the general sign in the second line of Eq.(\ref{pol22}).
%${\vec \varepsilon_2} \equiv \left({\vec e}_z - i {\vec e}_x\right)/\sqrt 2$
The corresponding curves are rotated by $\pi$ around the $Oz$-axis. Due to
the number of symmetry axis, this rotation is relevant only
for the atomic contribution and the differential cross section in Fig.6,
where $2\omega>|E_{1s}|$.

\section{CONCLUSIONS}

Our investigations show that the differential cross sections
for two-photon free-free transitions is strongly influenced by the
state of polarization of the two photons in the domain where the
dressing effects are important, that means for small scattering angles and
in the vicinity of atomic resonances.

When at least one polarization vector is complex, one mirror symmetry is
broken in  the azimuthal angular distribution of the scattered electron
if the virtual transitions to continuum
of the bound electron are energetically allowed, i.e., $2\omega>|E_{1}|$.
We have shown also that the differential cross sections of two-photon
free-free transitions are sensible with respect to the helicity
of the polarization vectors. These effects are present only if the
atomic diagrams (Fig.1c) are included in the calculation.

\vspace*{1 cm}

{\bf Acknowledgement}

\vspace*{0.5 cm}

This research is supported in part by the EC PECO contract
no ERB CIPD CT940025.
Part of the computations were performed in the Computer Center of
the Quantum and Statistical Physics Group (Bucharest-M\u{a}gurele),
supported by
SOROS Foundation. One of the authors (A.C.) is indebted to A. Maquet
for interesting discussions. A critical reading of the manuscript by
V. Florescu is warmly acknowledged.

\newpage
\vspace*{1cm}

{\bf Figure captions}\\
\begin{itemize}

\item {\bf Figure 1}
Feynman diagrams for two-photon processes in laser-assisted electron-atom
scattering. (a) electronic diagrams, (b) mixed diagrams,
and (c) atomic diagrams. The projectile is represented by a single line, the
bound electron by a double line.

\item {\bf Figure 2}
The differential cross section for two-photon absorption,
normalized with respect to field intensities $I_1I_2$,
as a function of the frequency of the photons.
$I_1$, $I_2$ and $d\sigma^{(+2)}/d\Omega$ are in a.u.
The energy of the projectile is $E_i=100$ eV; ${\vec k}_i || Oz$ and
the scattering angle is $\theta=5^0$.
The polarization vectors are linear:
${\vec \varepsilon}_1 || Oz$ and ${\vec \varepsilon}_2 || Ox $.
Solid line represents the differential cross section in Eq.(\ref{sec}),
dotted line the electronic contribution and
dot-dashed line the mixed contribution.

\item {\bf Figure 3}
The differential cross section for two-photon absorption,
normalized with respect to field intensities $I_1I_2$,
as a function of the azimuthal angle~$\phi$.
$I_1$, $I_2$ and $d\sigma^{(+2)}/d\Omega$ are in a.u.
The energy of the projectile is $E_i=100$~eV, ${\vec k}_i || Oz$
and the scattering angle is $\theta = 20^0$.
The frequency of the photons is $\omega=5$ eV,
${\vec \varepsilon_1} = {\vec e}_z$ and
${\vec \varepsilon_2} = \left({\vec e}_z +i {\vec e}_x\right)/\sqrt 2$.
The electronic, mixed and atomic contributions are also plotted in the
same conditions as the differential cross section.

\item {\bf Figure 4}
Same as Fig.3, but $\omega=10$ eV.

\item {\bf Figure 5}
Same as Fig.3, but
${\vec \varepsilon_1} \equiv \left({\vec e}_z +i {\vec e}_x\right)/\sqrt 2$
and
${\vec \varepsilon_2} \equiv \left({\vec e}_y +i {\vec e}_z\right)/\sqrt 2$.

\item {\bf Figure 6}
Same as Fig.5, but $\omega=10$ eV.

\end{itemize}


\newpage
\begin{thebibliography}{99}
\bibitem{exp} B. Wallbank and J. K. Holmes, J. Phys. B {\bf 27}, 1221
	      (1994), and J. Phys. B {\bf 27}, 5405(1994)

\bibitem{teo} A. Cionga, L. Dimou, and F.H.M. Faisal,
	      J. Phys. B {\bf 30}, L361 (1997); see also the references herein

\bibitem{bri} G. Kracke, J. S. Briggs, A. Dubois, A. Maquet,
		and V. V{\'e}niard, J. Phys. B {\bf 27}, 3241 (1994)

\bibitem{mar} M. D{\"o}rr, C. J. Joachain, R. M. Potvliege and
		S. Vu\u{c}i{\'c}, Phys. Rev. A {\bf 49}, 4852 (1994)

\bibitem{b-f} F.V. Bunkin and M.V. Fedorov,
		Zh. Eksp. Teor. Fiz. {\bf 49}, 1215 (1965)
                [Sov. Phys. JETP {\bf 22},884 (1966)]

\bibitem{vf1} V. Florescu and T. Marian, Phys. Rev. A {\bf 34}, 4641 (1986)

\bibitem{ac1} A. Cionga and V. Florescu, Phys. Rev. A {\bf 45}, 5282 (1992)

\bibitem{vf2} V. Florescu, A. Halasz, and M. Marinescu,
		Phys. Rev. A {\bf 47}, 394 (1993)

\bibitem{a-g} A. Cionga and G. Buic\u{a}, to be published

\end{thebibliography}
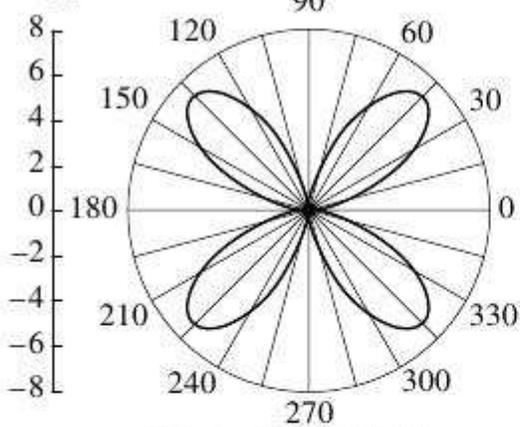
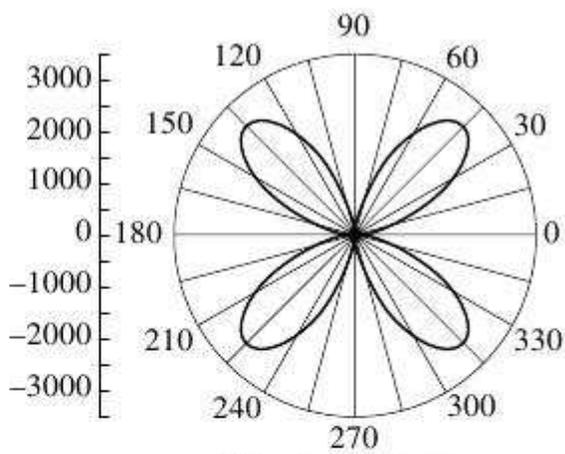
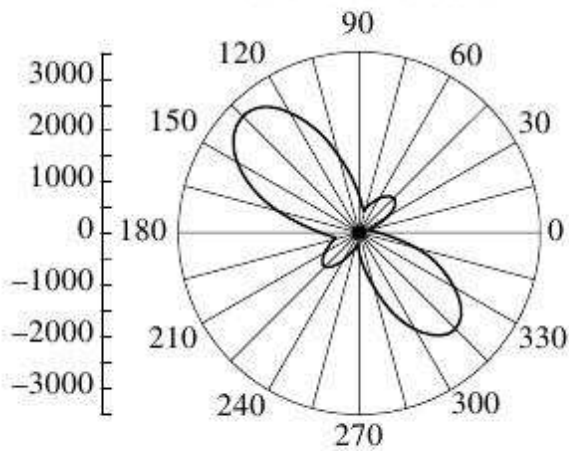
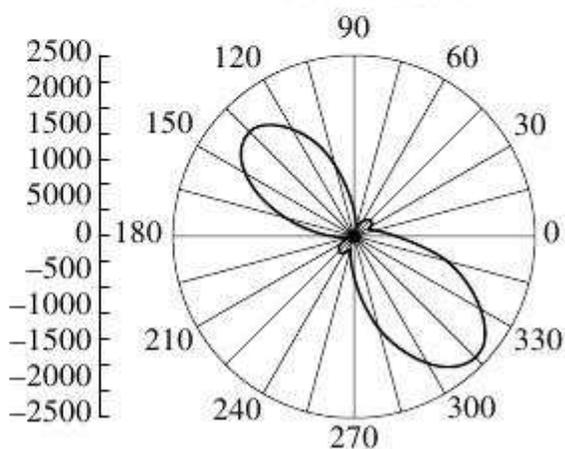
\end{document}